\documentclass[floatfix,prl,twocolumn,10pt]{revtex4}
\usepackage{amsmath}
\usepackage{amssymb}
\usepackage{times}
\usepackage{latexsym}
\usepackage{graphicx}
\usepackage[english]{babel}
\usepackage{verbatim}
\usepackage{bbm}
\usepackage[usenames,dvipsnames]{color}
\usepackage{hyperref}
\usepackage{tikz}

\begin{document}

\title{Ultimate communication capacity of quantum optical channels \\by solving the Gaussian minimum-entropy conjecture}
\author{V. Giovannetti$^{1}$, R. Garc{\'i}a-Patr\'on$^{2,3}$,
N. J. Cerf$^{2}$,  and A. S. Holevo$^{4}$}
\affiliation{$^{1}$NEST, Scuola Normale Superiore and Istituto Nanoscienze-CNR, I-56127 Pisa, Italy\\
$^{2}$QuIC,  Ecole Polytechnique de Bruxelles,  CP 165, Universit\'e Libre de Bruxelles, 1050 Bruxelles, Belgium  \\
 $^{3}$Max-Planck Institut f\"{u}r Quantenoptik, Hans-Kopfermann Str. 1, D-85748 Garching, Germany \\
 $^4$Steklov Mathematical Institute, RAS, Moscow}
\date{\today}
%
%
%
%\pacs{}
%
%
\begin{abstract}
Optical channels, such as fibers or free-space links, are ubiquitous in today's telecommunication networks. They rely on the electromagnetic field associated with photons to carry information from one point to another in space. As a result, a complete physical model of these channels must necessarily take quantum effects into account in order to determine their ultimate performances. Specifically, Gaussian photonic (or bosonic) quantum channels have been extensively studied over the past decades given their importance for practical purposes. In spite of this, a longstanding conjecture on the optimality of Gaussian encodings has yet prevented finding their communication capacity. Here, this conjecture is solved by proving that the vacuum state achieves the minimum output entropy of a generic Gaussian bosonic channel. This establishes the ultimate achievable bit rate under an energy constraint, as well as the long awaited proof that the single-letter classical capacity of these channels is additive. Beyond capacities, it also has broad consequences in quantum information sciences.

\end{abstract}
\maketitle

Since the advent of lasers, the question of how quantum effects must be accounted for in optical communication systems has been around~\cite{Gordon}. The limits imposed by quantum mechanics on the highest possible bit rate achievable through optical channels (the so-called {\it classical capacity}) have long been studied, see e.g.~\cite{CAVES}, but no complete resolution of the problem was known due to the extreme difficulty of identifying the optimal encodings and decodings. Recent progresses in quantum communication theory then moved this question forward by providing a formal expression of the classical capacity of quantum channels \cite{HOLEVO98,SCHWEST,BENSHOR}. Later on, the capacity of a large class of realistic quantum communication models known as bosonic Gaussian Channels (BGC) was expressed in \cite{HOWE}, by conjecturing the optimality of Gaussian encodings. These are most channels
using the electromagnetic field (or any bosonic field) as information carriers, e.g. dissipative optical fibers, optical waveguides, or free-space communication lines.

The Gaussian conjecture has consequently become one of the most debated conjecture in quantum communication theory over the last decade. It can be rephrased stating
that the minimum von Neumann entropy at the output of a BGC is  achieved by Gaussian input states (it is named {\it minimal output entropy} or {\it min-entropy} conjecture)~\cite{conj1}.
In its simplest form, it implies that a Bosonic system (say a mode of the e.m. field) interacting with a Gibbs thermal state via a quadratic exchange Hamiltonian
will attain the minimum possible entropy value if initially prepared into a coherent state, say the vacuum.
This apparently innocuous statement  turns out to have profound physical and technological implications~\cite{natphot,conj3,konig1,konig2,gp,WEEDB,conj2,QCMC}.
Indirect evidences of its validity have been accumulated
over the years~\cite{renyi1,renyi2,serafini,conj2,GUHA,gp,WEEDB,conj3,konig1,konig2,natphot,schaffer,QCMC}, yet its proof has remained elusive.

Here
we prove this Gaussian min-entropy conjecture for single-mode BGCs (a set including for instance thermal, additive classical noise, and amplifier channels~\cite{HOWE}), which in turn provides the ultimate closed expression for the communication capacity of BGCs as prescribed by quantum mechanics (an extended version of this theorem
which applies to multi-mode scenarios is given in~\cite{PROOF}.)
The proof restricts the entropy minimization over bounded energy states, but this does not affect the capacities. It also implies that the capacity of BGCs can be achieved via Gaussian encodings and is additive, which was a longstanding open question.  Furthermore, as sketched in the discussion, proving the conjecture has broader consequences, allowing us to compute the
entanglement of formation~\cite{EOF} for some non-symmetric Gaussian states or to deduce the
optimality of Gaussian measurements in the quantum discord~\cite{MODI} for Gaussian states~\cite{PIRA}. Other major implications can be anticipated.

   %%%%%%%%%%%%%%%%%%%%%%%%%%%%%%%%%%%%%%

\begin{figure}[t]
	\begin{center}
	\includegraphics[trim=0pt 0pt 0pt 0pt, clip, width=0.5\textwidth]{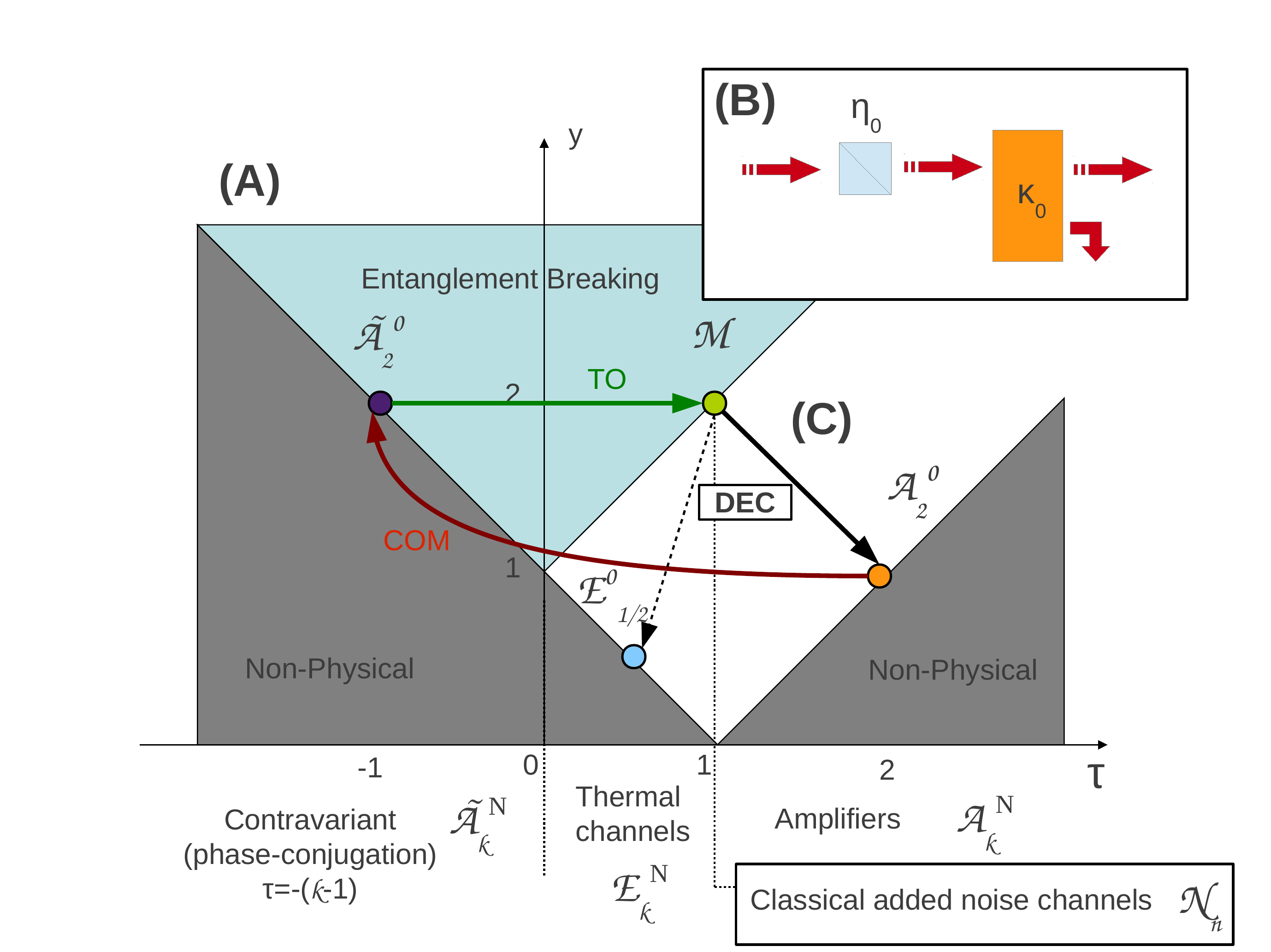}
\vspace{5pt}
		\caption{(Color online)
		\textbf{(A)} Single-mode phase-covariant (or contravariant) Gaussian Bosonic channels
		belong to four families characterized by the loss/gain parameter $\tau$:
		(i) the thermal channel ${\cal E}_\tau^{N}$ with $0\leq \tau\leq 1$;
		(ii) the amplifier channel ${\cal A}_\tau^{N}$ where $\tau\geq 1$;
		(iii)  the additive classical noise channel ${\cal N}_y$ satisfying $\tau=1$;
		(iv) the (contravariant) phase-conjugation channel with $\tau\leq 0$.
		All physical channels must satisfy $y\geq|\tau-1|$,
		and they are quantum-limited when this inequality is saturated.
		The channels where $y\geq|\tau|+1$ can be shown to be entanglement-breaking.	
		\textbf{(B)} A generic covariant single-mode channel $\Phi$, of parameters
		$\tau$ and $y$ can be expressed as the concatenation of a quantum-limited lossy channel
		followed by a quantum-limited amplifier $\Phi = {\cal A}_{\kappa_0} \circ {\cal E}_{\eta_0}$, where $\tau=\kappa_0\eta_0$ and $y=\kappa_0(1-\eta_0)+(\kappa_0-1)$.
		\textbf{(C)} Our proof relies on the following observation, which is applied iteratively. Any quantum-limited amplifier
		${\cal A}_{\kappa_0}$ has a complementary (COM) channel $\tilde{\cal A}_{\kappa_0}$ that is phase-conjugating
		(i.e. $\tau\leq 0$), hence entanglement-breaking.
		Applying a transposition operation (TO) to $\tilde{\cal A}_{\kappa_0}$ results in a new channel $\cal M$ satisfying
		$\tau=\kappa_0-1$ and $y=\kappa_0$, which can be decomposed (DEC) as
		${\cal A}_{\kappa_0}\circ {\cal E}_{(\kappa_0-1)/\kappa_0}$.}
		\label{channels-representation}
	\end{center}
\end{figure}

\paragraph{The Gaussian channel model:--}

BGCs are the quantum counterparts of the Gaussian channels of classical information theory~\cite{COVER}.
They describe all physical completely-positive trace-preserving (CPT) transformations~\cite{h}  which, when acting on a collection of Bosonic degrees of freedom (continuous-variable
quantum systems~\cite{BRAU,CERFBOOK}) preserve their Gaussian
character~\cite{HOWE}. Accordingly, BGCs represent the most common noise models which tamper quantum optical implementations,
including those responsible for the attenuation, amplification, and squeezing of optical signals~\cite{CAVES}.
The most physically relevant cases are the phase-covariant or contravariant BGCs operating on a single Bosonic mode described by the annihilation
(resp. creation) operator $a$ (resp. $a^\dag$), fulfilling canonical commutation rules $[a,a^\dag]=1$.
A general single-mode covariant GBC transforms the symmetrically-ordered
characteristic function
$\chi(z)={\rm Tr}[\rho D(z)]$  ($z$ being complex) associated with the
input state $\rho$~\cite{walls} as
$\chi(z) \rightarrow \chi(\sqrt{\tau} z)\exp(-y|z|^2/2)$,  where $\tau\ge
0$ is a loss/gain parameter
and $y$ parametrizes the added noise, while $D(z)=\exp[z a^\dag - z^* a]$
is the displacement operator.
For contravariant (phase-conjugation) channels, we
have $\chi(z) \rightarrow \chi(-\sqrt{|\tau|} z^*)\exp(-y|z|^2/2)$, where
$\tau\leq 0$.
The GBC is physical (the map is CPT) provided $y$ satisfies $y\geq|\tau-1|$.
The map is called quantum-limited when this latter inequality is saturated.
A compact representation of phase-covariant (contravariant) BGC channels \cite{schaffer} is given in
Figure~\ref{channels-representation}.
Among these channels, one identifies four fundamental classes (adopting the notation of  Refs.~\cite{conj1,conj2,renyi1,renyi2,natphot,CARV}):
the thermal channels ${\cal E}_\eta^N$, the amplifier channels ${\cal A}_\kappa^N$ and their weak conjugate~\cite{CARV,cgh,hoc}
$\tilde{\cal A}_\kappa^N$, and the classical additive noise channels ${\cal N}_n$
(see Fig.~\ref{channels-representation} and Supplemental Information for a precise definition).
The channel ${\cal E}_\eta^N$  can be effectively
described as the transformation induced by mixing on  a beam-splitter of transmissivity $\eta\in [0,1]$
 the input state  of the system  with an external mode $B$ initialized into a Gibbs thermal Bosonic state $\rho_G^{(N)} =\left(\tfrac{N}{N+1}\right)^{b^\dag b}/(N+1)$ with
 average photon number $N\in [0,\infty[$.
The quantum-limited element of the set ${\cal E}_\eta^N$ corresponds to $N=0$, a purely lossy (attenuator) channel
${\cal E}_\eta:={\cal E}_\eta^{0}$~\cite{gio}.
A well-know property of this channel is that, when applied iteratively on a state, it brings it toward  the vacuum state, i.e.
\begin{eqnarray}  \label{mixing}
\lim_{q\rightarrow \infty} [{\cal E}_\eta]^q (\rho) = |0\rangle\langle 0|\;.
\end{eqnarray}
The amplifier channels  ${\cal A}_\kappa^N$ and their weak-conjugate $\tilde{\cal A}_\kappa^N$
refer, respectively, to the signal and idler modes one gets at the output of a parametric amplifier
that couples (via a two-mode squeezing transformation)  the input state with the $B$ mode defined above.
These channels are characterized by the gain parameter $\kappa\in [1,\infty[$ and for $N=0$ represent
quantum-limited transformations (${\cal A}_\kappa:={\cal A}^{0}_\kappa$ and
$\tilde{\cal A}_\kappa:=\tilde{\cal A}^{0}_\kappa$ respectively). Finally, the classical additive noise channel ${\cal N}_n$,
is induced by randomly displacing the input states in phase space via a Gaussian probability distribution of
variance $n\in [0,\infty[$. Note that ${\cal E}_\eta^N$, ${\cal A}_\kappa^N$, and ${\cal N}_n$ are covariant, while $\tilde{\cal A}_\kappa^N$ is contravariant
under a phase shift of the input state.

 \paragraph{The Gaussian channel conjecture:--}

Consider a single-mode BGC effecting the map $\Phi$. Following the prescription detailed in Refs.~\cite{HOLEVO98,SCHWEST,h2,HOSHI,h}
its optimal classical communication rate, or classical capacity~\cite{BENSHOR}, can be computed as
$C(\Phi )=\lim_{m\rightarrow \infty }\ \frac{1}{m}C_{\chi }(\Phi ^{\otimes
m})$
where $\Phi ^{\otimes m}$ describes $m$ channel uses (memoryless
noise model), while $C_{\chi }[\Psi]$ is the single-letter or $\chi$-capacity of the map $\Psi$ defined by the
expression
$C_{\chi }(\Psi )=\max_{\mbox{{\tiny ENS}}}\left\{ S(\Psi(\rho_
{\mbox{{\tiny ENS}}})) -\sum_{j}p_{j}S(\Psi \lbrack \rho _{j}])\right\}$.
Here, the maximization is performed over the set of
input ensembles $\mbox{ENS}=\{p_{j};\rho _{j}\}$ ($p_{j}$ being
probabilities, while $\rho _{j}$ being density matrices),
where $\rho_{\mbox{{\tiny ENS}}} = \sum_j p_j \rho_j$ is the associated
 average state, and $S(\rho )=-\mbox{Tr}[\rho \log_2 \rho ]$ is the von Neumann entropy of $\rho$~\cite{h}.
The input ensembles $\mbox{ENS}$ must respect an {\it average} input energy constraint, i.e.,
$\mbox{Tr} [H^{(m)} \rho_{\mbox{\tiny ENS}}] \leq m E$,
where $H^{(m)}=\sum_{j=1}^m a^\dag_j a_j$ is the total photon number operator in the $m$ modes.
We also introduce the quantity
$S_{\min}^{(<)}[\Phi^{\otimes m}] := \min_\psi S(\Phi^{\otimes m} (|\psi\rangle \langle \psi|))$
for the map $\Phi^{\otimes m}$, where the symbol $^{(<)}$ indicates that the minimization is restricted over $m$-mode states $|\psi\rangle$ having bounded mean input  energy, i.e.,
$\mbox{Tr} [H^{(m)} |\psi\rangle \langle \psi|] <  \infty$.  This restriction ensures finiteness of all the entropy quantities throughout the paper.

\begin{figure}[t]
	\begin{center}
	\includegraphics[trim=0pt 0pt 0pt 0pt, clip, width=0.5\textwidth]{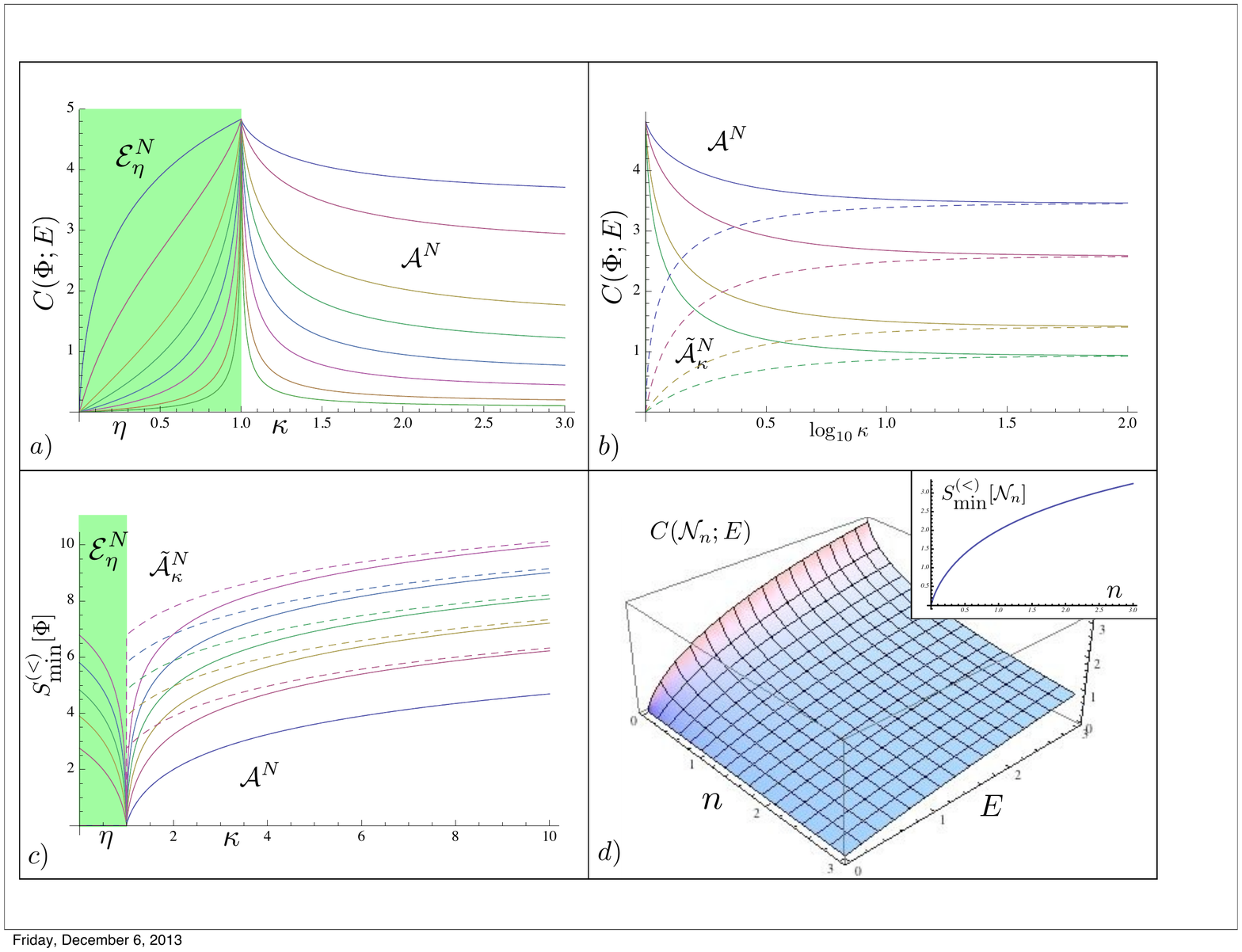}
\vspace{5pt}
		\caption{(Color online) Plots of the capacities and of the minimal output entropies for thermal
		${\cal E}_\eta^N$, classical additive noise ${\cal N}_n$,
		amplifier ${\cal A}_\kappa^N$,
  contravariant amplifier $\tilde{\cal A}_\kappa^N$ BGCs.
   The exact definition of the channels and the analytical expressions of these quantities are given in the Supplemental Information.
  a) Plot of the capacity of
  ${\cal E}_\eta^N$ (green region)
		and  ${\cal A}_\kappa^N$ as a function of the parameters
		$\eta$ and $\kappa$ respectively for different values of $N$
		(from top to bottom $N=0, 1, 5,10,20,40,100$, and $200$). For ${\cal E}_\eta^N$ the capacity of $N=0$ coincides with the one given in
		\cite{gio}.
		b) Comparison between the capacity of ${\cal A}_\kappa^N$ (continuous lines) and $\tilde{\cal A}_\kappa^N$ (dashed lines) as a function of $\kappa$ for $N=0,1,5,10$ (from top to bottom).  c) minimum output entropies for
		${\cal E}_\eta^N$ (green region), ${\cal A}_\kappa^N$ (continuos lines), and $\tilde{\cal A}_\kappa^N$ (dashed lines). From
		 bottom to top $N=0,2,5.10,20,40$. d) Plot of the capacity of ${\cal N}_n$ as a function of the noise parameter $n$ and of the energy constraint parameter $E$.  Inset, minimum output entropy of ${\cal N}_n$. In all plots the plots $E = 10$.}\label{fig1}
	\end{center}
\end{figure}

The min-entropy conjecture then implies that $S_{\min}^{(<)}[\Phi^{\otimes m}]$ should be additive and equal to $m$ times the output entropy
associated with the vacuum input $|0\rangle$, i.e.
\begin{eqnarray}
 \label{minentropy}
S_{\min}^{(<)}[\Phi^{\otimes m}]  =  m\;  S_{\min}^{(<)}[\Phi]
  = m\;  S(\Phi(|0\rangle\langle 0|))\;.
\end{eqnarray}
While avoiding technicalities associated with the continuity of the entropy functional~\cite{WEHRL},
restricting the minimization over bounded-energy states turns out to be sufficient to derive several important results.  In particular,
as detailed in the Supplemental Information, the validity of~(\ref{minentropy}) allows one to show
a couple of  properties first conjectured in~Ref.~\cite{HOWE} and  known to hold at least  for purely lossy maps~${\cal E}_\eta:={\cal E}_\eta^{0}$~\cite{gio}.
First, the fact that  Gaussian states provide optimal ensembles ENS for the classical capacity $C(\Phi; E)$ under average input energy constraint;
secondly, the fact that the  $C_\chi$ is additive,  i.e., $C(\Phi;E)= C_\chi(\Phi;E)$. The minimum entropy values $S_{\min}^{(<)}$ as well as the corresponding capacities $C$
for the four classes of Gaussian channels are provided in the Supplemental Information.  Plots of these functions are presented in Figure~\ref{fig1}.

%%%%%%%%%%%%%%%%
%%%%%%%%%%%%%%%%%%%

\paragraph{The proof:--}

Following Refs.~\cite{gp},
 we use the fact that a generic  phase covariant (resp. contravariant)  single-mode channel $\Phi$ can be expressed as
the concatenation of a quantum-limited lossy channel followed
by a quantum-limited amplifier (resp. quantum-limited contravariant amplifier). Thus,
\begin{eqnarray}
\Phi =
 {\cal A}_{\kappa_0} \circ {\cal E}_{\eta_0} \quad \mbox{(or~}  \Phi = {\tilde{\cal A}}_{\kappa_0} \circ {\cal E}_{\eta_0}  ) \;,  \label{dec111}
\end{eqnarray}
where the parameters $\kappa_0$ and $\eta_0$ are in biunivocal correspondence with the selected channel $\Phi$
and are obtained by solving the equations $\tau=\eta_0\kappa_0$ and $y=\kappa_0(1-\eta_0)+(\kappa_0-1)$ (see Supplemental Information).
  Since channels ${\cal E}_{\eta_0}$  map
  the vacuum into the vacuum, it follows that to prove the min-entropy
  conjecture for one use of $\Phi$,
  it is sufficient to prove it for ${\cal A}_{\kappa_0}$, or equivalently for
  $\tilde{\cal A}_{\kappa_0}$. These last two
  have indeed the same minimum output entropy as they are
  conjugate~\cite{h1,mkr} (more generally, given a generic pure input state $|\psi\rangle$, the density operators
   ${\cal A}_{\kappa_0}(|\psi\rangle\langle \psi|)$ and $\tilde{\cal A}_{\kappa_0}(|\psi\rangle\langle \psi|)$
   have the same  nonzero spectra.) Applying this single-mode reduction to $m$ channel uses,
   $\Phi^{\otimes m} = {\cal A}_{\kappa_0}^{\otimes m}  \circ {\cal E}_{\eta_0}^{\otimes m} $,
   we actually need to prove the min-entropy conjecture for channel ${\cal A}_{\kappa_0}^{\otimes m} $ since the product vacuum state is invariant under
   ${\cal E}_{\eta_0}^{\otimes m}$. A further simplification then follows
   by noting that the conjugate channel $\tilde{\cal A}_{\kappa_0}$ is entanglement-breaking~\cite{hoEB}. This implies that its minimal output entropy is additive,
   $S^{(<)}_{\min}[\tilde{\cal A}_{\kappa_0}^{\otimes m}] = m \, S^{(<)}_{\min}[\tilde{\cal A}_{\kappa_0}]$, so that the same is true for ${\cal A}_{\kappa_0}$.
    Consequently, the minimum output entropy for $m$ uses of the channel $\Phi$ can always be achieved over separable inputs, and the additivity of $C_{\chi }$ holds provided we can prove the single-mode amplifier conjecture~\cite{QCMC}
 \begin{eqnarray}
S^{(<)}_{\min}[{\cal A}_{\kappa_0}] =  S({\cal A}_{\kappa_0}(|0\rangle\langle 0|))\;. \label{minentropy1}
\end{eqnarray}

\begin{figure}[t]
	\begin{center}
	\includegraphics[trim=0pt 0pt 0pt 0pt, clip, width=0.4\textwidth]{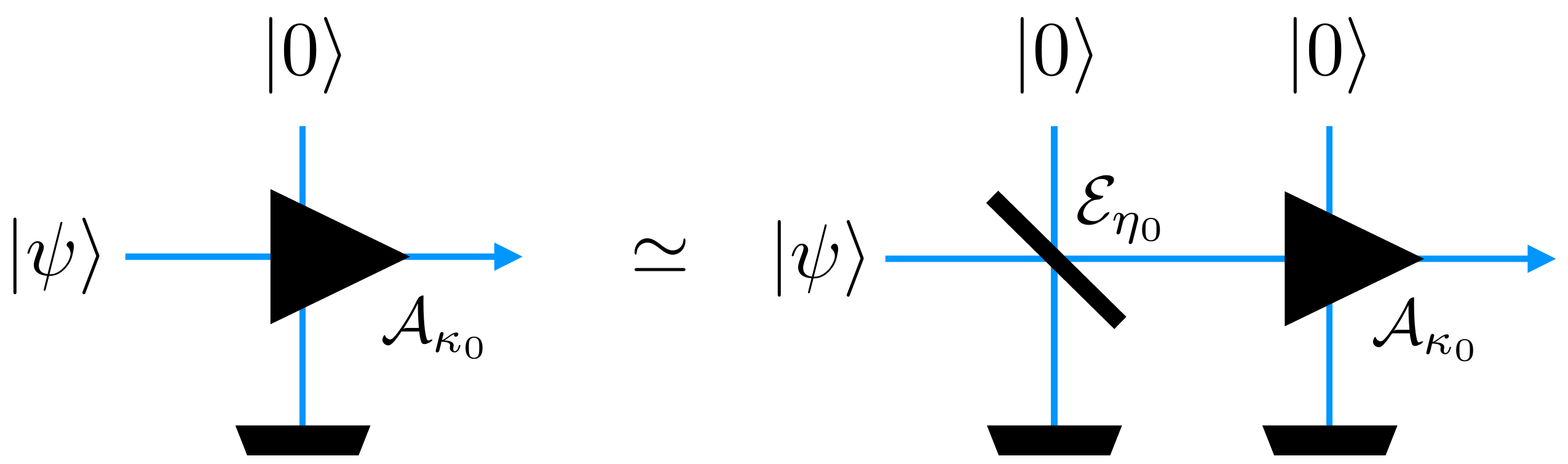}
\vspace{5pt}
		\caption{(Color online) Graphical representation of  Eq.~(\ref{eq8}).
Here the symbol ``$\simeq$" indicates that  applying
 the quantum-limited amplifier ${\cal A}_{\kappa_0}$ to the pure input $|\psi\rangle$ or to its evolved counterpart  via a
 lossy  channel of transmissivity $\eta_0 = 1 - 1/\kappa_0$  (i.e.  ${\cal E}_{\eta_0}(|\psi\rangle\langle \psi|)$) produces
 outputs which have the same spectra.	 }\label{fig2}
	\end{center}
\end{figure}

To prove it,  we observe that the conjugate channel $\tilde{\cal A}_{\kappa_0}$
     can be expressed as the following measure-and-prepare transformation in the coherent state basis
  \begin{equation}
\rho \mapsto \tilde{\cal A}_{\kappa_0}  [\rho]=\int \frac{d^2 z}{\pi}|-\sqrt{\kappa_0}{z}^*
\rangle \langle -\sqrt{\kappa}_0{z}^*|\;{\langle }z|\rho | z\rangle \;, \label{EQ1}
\end{equation}
where $|z\rangle=D(z) |0\rangle$ is a coherent state ($z$ being complex).
Next step is to notice that by taking the complex conjugate of $\tilde{\cal A}_{\kappa_0}  [\rho]$
with respect to the Fock basis, one gets a state that coincides with the output state of
an entanglement-breaking covariant channel ${\cal M}$
of parameters $\tau=\kappa_0-1$ and $y=\kappa_0$ (see Fig.~\ref{channels-representation}).
Thus $\tilde{\cal A}_{\kappa_0}  [\rho]$ and ${\cal M}[\rho]$
share the same spectrum for a fixed input $\rho$.
%the amplifier channel characterized by
%gain parameter $\kappa_0$ and $N=\kappa_0$, i.e.   ${\cal A}_{\kappa_0}^{\kappa_0}$.
%Thus $\tilde{\cal A}_{\kappa_0}  [\rho]$ and ${\cal A}_{\kappa_0}^{\kappa_0}[\rho]$
%share the same spectrum for a fixed input $\rho$.
Together with the fact that for a pure input state $|\psi\rangle$, $\tilde{\cal A}_{\kappa_0}(|\psi\rangle\langle \psi|)$
also has the same spectrum as ${\cal A}_{\kappa_0}(|\psi\rangle\langle \psi|)$, this implies that there exists a unitary
transformation ${\cal U}$ (possibly dependent upon $|\psi\rangle$) which fulfills the condition
\begin{eqnarray}
{\cal A}_{\kappa_0} ( | \psi\rangle\langle \psi|) &=& [{\cal U}\circ {\cal M} ](|\psi\rangle\langle \psi|)
\nonumber \\
&=& [{\cal U}\circ {\cal A}_{\kappa_0}\circ {\cal E}_{\eta_0} ](|\psi\rangle\langle \psi|)\;, \label{eq8}
 \label{ff}
\end{eqnarray}
where the second identity follows from~Eq.~(\ref{dec111}) which allows
one  to write ${\cal M}={\cal A}_{\kappa_0}\circ {\cal E}_{\eta_0}$
with $\eta_0 =({\kappa_0-1})/{\kappa_0}$ (see Fig.~\ref{channels-representation}).
Equation~(\ref{ff}) means that for a pure input state $|\psi\rangle$, the output state of a quantum-limited amplifier ${\cal A}_{\kappa_0}$ coincides
(up to a unitary transformation which in principle depends upon $|\psi\rangle$)
 with the one obtained by first applying a lossy channel ${\cal E}_{\eta_0}$
 to the input and then ${\cal A}_{\kappa_0}$, as shown in Fig.~\ref{fig2}.
 Consider next a generic ensemble decomposition of ${\cal E}_{\eta_0}(|\psi\rangle\langle \psi|)$, i.e.
 ${\cal E}_{\eta_0}
 (|\psi\rangle\langle \psi|) = \sum_j p_j |\psi_j\rangle\langle \psi_j|$
with $p_j>0$ being probabilities. Replacing this into Eq.~(\ref{eq8})
and iterating the same passage $q$ times, we get the identity
\begin{eqnarray}
 {\cal A}_{\kappa_0} ( | \psi\rangle\langle \psi|) &=&  \sum_\ell q_\ell \;  [{\cal W}_\ell \circ  {\cal A}_{\kappa_0}](|\phi_\ell\rangle\langle \phi_\ell|)
 \;, \label{eq8910}
\end{eqnarray}
where ${\cal W}_\ell$ are unitaries,  $q_\ell>0$ are probabilities,  and $|\phi_\ell\rangle$ are state vectors which provide an ensemble decomposition
after $q$ applications of  ${\cal E}_{\eta_0}$ on the input state $|\psi\rangle$, i.e.
\begin{eqnarray}
\sum_\ell  q_\ell |\phi_\ell\rangle\langle \phi_\ell| = [{\cal E}_{\eta_0}]^q
 (|\psi\rangle\langle \psi|) \;.
\label{cond1}
 \end{eqnarray}
 (Here, the index $\ell$ actually refers to a path of $j$ indices across the $q$ subsequent ensemble decompositions of the output state of ${\cal E}_{\eta_0}$ -- see Ref.~\cite{PROOF} for details)
%%%%%%%%%%%%%%%%%%%%%%%%%%%%%%%%%%%
%%%%%%%%%%%%%%%%%%%%%%%%%%%%%%%%%%%%
Exploiting the concavity of the von Neumann entropy, the identity~(\ref{eq8910}) yields
\begin{eqnarray}
S( {\cal A}_{\kappa_0} ( | \psi\rangle\langle \psi|)) &\ge&  \sum_\ell q_\ell \;  S( {\cal A}_{\kappa_0}(|\phi_\ell\rangle\langle \phi_\ell|) )
 \;. \label{eq89101}
\end{eqnarray}
Reminding property~(\ref{mixing}), it is intuitively clear at this level that for $q\rightarrow \infty$, the ensemble $\{q_\ell, |\phi_\ell\rangle\}$
in Eq.~(\ref{cond1}) will contain one single component, namely the vacuum.
Then, Eq.~(\ref{eq89101}) will lead us conclude that the output entropy for input $|\psi\rangle$ is larger than the one associated
with the vacuum, hence to Eq.~(\ref{minentropy1}).
To show this rigorously, we start from the inequality
\begin{eqnarray}
S({\cal A}_{\kappa_0} (|\phi _{\ell}\rangle \langle \phi _{\ell}|))&\geq& -\mbox{Tr}[{\cal A}_{\kappa_0} (|\phi
_{\ell}\rangle \langle \phi _{\ell}|)\log_2 {\cal A}_{\kappa_0} (\sigma)]
\nonumber \\
&&+\mbox{Tr}[|\phi
_{\ell}\rangle \langle \phi _{\ell}|\log_2 \sigma] ,
\end{eqnarray}
which holds for an arbitrary state $|\phi_\ell\rangle$ of Eq.~(\ref{cond1})
and for a generic state $\sigma$, and
which can be easily derived from the monotonic decrease~\cite{h}
of the relative entropy $S(|\phi_\ell\rangle\langle \phi_\ell \| \sigma)$ under the action of the map ${\cal A}_{\kappa_0}$. Setting $\sigma = [{\cal E}_{\eta_0}]^q(|\psi\rangle\langle \psi|)$ and replacing into the rhs of Eq.~(\ref{eq89101}) we then get
\begin{eqnarray}
S( {\cal A}_{\kappa_0} ( | \psi\rangle\langle \psi|))
&\geq& S( ({\cal A}_{\kappa_0} \circ [{\cal E}_{\eta_0}]^q)( | \psi\rangle\langle \psi|)) \nonumber \\
&& -  S(  [{\cal E}_{\eta_0}]^q( | \psi\rangle\langle \psi|))\;. \label{almost}
\end{eqnarray}
Since $|\psi\rangle$ is of bounded mean energy,
the same holds for $[{\cal E}_{\eta_0}]^q( | \psi\rangle\langle \psi|)$
and $({\cal A}_{\kappa_0} \circ [{\cal E}_{\eta_0}]^q)( | \psi\rangle\langle \psi|)$ [the corresponding  energy expectations values being in fact
$\eta^q \langle \psi| a^\dag a |\psi\rangle$ and
$\kappa \eta^q \langle \psi| a^\dag a |\psi\rangle + \kappa (\kappa -1)$].
We can then invoke the continuity of the von Neumann entropy \cite{WEHRL} and use
property (\ref{mixing}) for $q\rightarrow \infty$ to  show that
\begin{eqnarray}
S( {\cal A}_{\kappa_0} ( | \psi\rangle\langle \psi|))
&\geq&   S(  {\cal A}_{\kappa_0} ( | 0 \rangle\langle 0 |))\;, \label{almost1}
\end{eqnarray}
and hence Eq.~(\ref{minentropy1}). This proves the conjecture (\ref{minentropy}).

\paragraph{Discussion:--}

Aside from a definitive proof of the classical capacity of the four fundamental Gaussian channels ($\Phi={\cal E}_\eta^N$, ${\cal N}_n$, ${\cal A}_\kappa^N$ or $\tilde{\cal A}_\kappa^N$),
the above approach can be used to prove an identity analogous to (\ref{minentropy}) but for arbitrary (not necessarily covariant or contravariant) single-mode channels $\Phi$. As explained in the Supplementary Information, the optimality of Gaussian inputs for $C(\Psi; E)$ and the additivity of $C_\chi(\Psi; E)$ hold, however, only for large enough values of the mean energy $E$ for such channels~\cite{LUPO}. The solution of the min-entropy conjecture also allows one to extend
the results of Refs.~\cite{ENTFORM1,ENTFORM} by yielding the exact formula for
the entanglement of formation (EoF)~\cite{EOF} of some non-permutation-symmetric two-mode Gaussian states.
(The EoF is a measure of the cost of generating a given quantum bipartite state $\rho$.)
Consider the two-mode density matrices $\rho(\kappa,N)$ one obtains at the output of a parametric
amplifier with gain parameter $\kappa$,
when injecting the vacuum in one port and a Gibbs thermal state
$\rho_G^{(N)}$ in the other, i.e.
$\rho(\kappa,N) = U_\kappa [ \rho_G^{(E)} \otimes |0\rangle\langle 0| ]U_\kappa^\dag$.
The symmetric case $\rho(\kappa,0)$ corresponds to the two-mode squeezed vacuum state, whose entanglement of formation is trivial to compute, i.e., $\mbox{EoF}[\rho(\kappa,0)] =g(\kappa -1)$.
For a general $\rho(\kappa,N)$ state, Eq.~(\ref{minentropy1}) implies
\begin{eqnarray} \label{eof1}
\mbox{EoF}[\rho(\kappa,N)] &=& \mbox{EoF}[\rho(\kappa,0)] = g(\kappa -1)\;,
\end{eqnarray}
(Notice that  this expression does not depend upon the mean photon number
$N$ of  $\rho_G^{(N)}$.)
The proof of this identity follows by first noticing that one can
obtain the upperbound $\mbox{EoF}[\rho(\kappa,N)]\leq g(\kappa -1)$ by generating $\rho(\kappa,N)$ starting from
the two-mode squeezed vacuum state  $\rho(\kappa,0)$
and  applying random correlated displacements (LOCC post-processing) to both  modes \cite{ENTFORM1}.
To show that this quantity is also a lower bound for
$\mbox{EoF}[\rho(\kappa,N)]$, one can use the fact that  the reduced density matrix of $\rho(\kappa,N)$ coincides with the output state ${\cal A}_\kappa(\rho_G^{(N)})$ of the quantum-limited amplifier with a thermal state
$\rho_G^{(N)}$ input. Exploiting the equivalence relation between the EoF and the minimum output entropy introduced in~\cite{WINTER}, our proof of  conjecture (\ref{minentropy}) implies Eq.~(\ref{eof1}) (see Supplementary Information for detailed proof).

 \textit{Acknowledgements}: The authors are grateful to L. Ambrosio, A. Mari, C. Navarrete-Benloch,
J. Oppenheim, M. E. Shirokov, R. F. Werner and A. Winter for comments and discussions. They also acknowledge
support and catalysing role of the Isaac Newton Institute for Mathematical
Sciences, Cambridge, UK: important part of this work was conducted when attending the Newton Institute programme \textit{Mathematical Challenges in Quantum Information}. AH acknowledges the Rothschild Distinguished Visiting Fellowship which enabled him to participate
in the programme and partial support from RAS Fundamental
Research Programs, Russian Quantum Center and RFBR grant No 12-01-00319.
N. J. C. and R. G.-P. also acknowledge financial support from the F.R.S.-FNRS under projects T.0199.13 and HIPERCOM, as well as from the Interuniversity Attraction Poles program of the Belgian Science Policy Office under Grant No. IAP P7-35 ¡Æ¡ÆPhotonics@be¡Ç¡Ç. R. G.-P. is Postdoctoral Researcher of the F.R.S.-FNRS.

\newpage

\section{Supplemental Information}

\subsection{Four fundamental Gaussian channels}

A compact way to represent Gaussian bosonic channels is obtained by expressing the density matrices $\rho$  of the mode as
$\rho = (1/\pi)\int d^2 z \chi(z) D(-z)$ where  the integral is performed over all $z$ complex
and where  $D(z):=\exp[z a^\dag - z^* a]$ is the displacement operator of the system while
 $\chi(z) := \mbox{Tr}[ \rho D(z)]$ is the  symmetrically ordered characteristic function of $\rho$~\cite{walls}.
Accordingly  the maps ${\cal E}_\eta^N$, ${\cal A}_\kappa^N$,  $\tilde{\cal A}_\kappa^N$ and ${\cal N}_n$ can be assigned
to the input-output  mappings $\chi(z) \rightarrow \chi'(\rho) = \mbox{Tr} [\Phi(\rho) D(z)]$
where~\cite{conj1,CARV}
\begin{eqnarray}
\chi'(z) = \left\{
\begin{array}{ll}
\chi(\sqrt{\eta} z) \; e^{-(1-\eta)(N+1/2) |z|^2}, & (\Phi={\cal E}_\eta^N)
\nonumber  \\
\chi(z) \; e^{-n |z|^2}\;,&
(\Phi={\cal N}_n)
\\
\chi(\sqrt{\kappa}z) \; e^{-(\kappa -1 )(N+1/2) |z|^2}\;,&
 (\Phi={\cal A}_\kappa^N)
\\
\chi(-\sqrt{\kappa-1}z^*) \; e^{-\kappa(N+1/2) |z|^2}\;,&
 (\Phi=\tilde{\cal A}_\kappa^N).
\end{array}
\right. \nonumber \\
\label{def}
\end{eqnarray}

\subsection{Gaussian Channel Decomposition}

As shown in Ref.~\cite{gp}, a generic phase-covariant single-mode channel
$\Phi_\tau^y$ of loss/gain parameter $\tau$ and added noise $y$
can be expressed as the concatenation of a quantum-limited lossy channel followed
by a quantum-limited amplifier $\Phi = {\cal A}_{\kappa_0} \circ {\cal E}_{\eta_0}$ (see Figure 1).
Respectively, for phase-contravariant single-mode channels, the decomposition reads
$\Phi = {\tilde{\cal A}}_{\kappa_0} \circ {\cal E}_{\eta_0})$. In both cases
the decomposition is unique and obtained from the equations
$\tau=\eta_0\kappa_0$ and $y=\kappa_0(1-\eta_0)+(\kappa_0-1)$.
Respectively, for phase-contravariant single-mode channels, the decomposition reads
$\Phi = {\tilde{\cal A}}_{\kappa_0} \circ {\cal E}_{\eta_0})$. In both cases
the decomposition is unique and obtained from solving the equations
$\tau=\eta_0(1-\kappa_0)$ and $y=(\kappa_0-1)(1-\eta_0)+\kappa_0$.
The explicit values of the parameter $\eta_0$ and $\kappa_0$ for the
maps ${\cal E}_\eta^N$, ${\cal A}_\kappa^N$,  $\tilde{\cal A}_\kappa^N$ and ${\cal N}_n$ are~\cite{gp,schaffer}
\begin{itemize}
\item Thermal channel ${\cal E}_\eta^N$:  $\eta_0=\eta/(1+(1-\eta)N)$, $\kappa_0=1+(1-\eta)N$;
\item Additive classical noise channel ${\cal N}_n$: $\eta_0=1/(n+1)$, $\kappa_0=n+1$ (see also~\cite{conj1});
\item Amplifier  channel ${\cal A}_\kappa^N$: $\eta_0=k/[k+(k-1)N]$, $\kappa_0=k+(k-1)N$.
\end{itemize}

\subsection{Linking the minimal entropy to the classical capacity}
\label{capandminentropy}

The $m$-mode energy-constrained $\chi$-capacity of a BGC $\Phi$ is expressed as
\begin{eqnarray}
C_{\chi }(\Phi^{\otimes m};E)&=&\max_{\mbox{{\tiny ENS}}}\big\{ S(\Phi^{\otimes m} \lbrack \sum_{j}p_{j}\rho
_{j}]) \nonumber \\
&&-\sum_{j}p_{j}S(\Phi^{\otimes m} \lbrack \rho _{j}])\big\},
\label{oneshotunassist10}
\end{eqnarray}
with the maximization being performed over all possible ensemble $
\mbox{ENS}=\{ p_j, \rho_j\}$ whose average state  $\rho=\sum_j p_j \rho_j$ belongs to the set ${\cal B}_E$ fulfilling the average energy constraint
\begin{eqnarray}
\mbox{Tr} [H^{(m)} \rho_{\mbox{\tiny ENS}}] \leq m E\;. \label{MINE}
\end{eqnarray}
where $H^{(m)}=\sum_{j=1}^m a^\dag_j a_j$ is the total photon number operator in the $m$ modes.

An upper-bound for~(\ref{oneshotunassist10}) can be easily obtained by replacing the first
term entering the maximization on the rhs with  the maximum output
entropy attainable within the set ${\cal B}_E$, i.e.
\begin{eqnarray}
S^{(E)}_{max}(\Phi^{\otimes m}) =\max_{\rho\in {\cal B}_E} S(\Phi^{\otimes m}(\rho))\;.
\end{eqnarray}
For the maps~(\ref{def}) this quantity is additive in $m$ and it is
attained by $[\rho_G^{(E)}]^{\otimes m}$ with   $\rho_G^{(E)}$ being a Gaussian state representing the Gibbs thermal state whose mean energy  coincides with $E$,
\begin{eqnarray}
S^{(E)}_{max}(\Phi^{\otimes m}) =m \, S^{(E)}_{max}(\Phi) = m \, S(\Phi(\rho_G^{(E)}))\;,
\end{eqnarray}
 (see e.g.~\cite{natphot,konig2,konig1,conj3}).
Accordingly, we can write
\begin{eqnarray}
C_{\chi }(\Phi^{\otimes m};E)&\leq &m \; S(\Phi(\rho_G^{(E)}))
- \min_{\mbox{{\tiny ENS}}} \sum_{j}p_{j} S(\Phi^{\otimes m} \lbrack \rho _{j}]) \nonumber\\
&&
\label{oneshotunassist11}
\end{eqnarray}

In contrast with the conventional version of the min-entropy conjecture~\cite{conj1},
we define here the min-entropy quantity
\begin{eqnarray}
S_{\min}^{(<)}[\Phi^{\otimes m}] := \min_{\rho \in {\cal B}^{(<)}} S(\Phi^{\otimes m} (|\psi\rangle \langle \psi|))  \, ,
\end{eqnarray}
associated with the map $\Phi^{\otimes m}$,
where the minimization is now restricted over the set ${\cal B}^{(<)}$ of $m$-mode states $\rho$ having bounded mean input  energy, i.e.,
$\mbox{Tr} [H^{(m)} \rho] <  \infty$. This allows us to rewrite the upper bound as
\begin{eqnarray}
C_{\chi }(\Phi^{\otimes m};E)&\leq &m \;S(\Phi(\rho_G^{(E)}))
- S_{\min}^{(<)}[\Phi^{\otimes m}] \;,
\label{oneshotunassist11bis}
\end{eqnarray}
where the last inequality follows by exploiting the fact
that for all $\rho_j$ entering in one of the
allowed  ensembles $\mbox{ENS}=\{ p_j, \rho_j\}$, the term $S(\Phi^{\otimes m} \lbrack \rho _{j}])$ can be lower bounded by
$S_{\min}^{(<)}[\Phi^{\otimes m}]$. This can be seen by noticing that
in order  to satisfy Eq.~(\ref{MINE}) all states $\rho_j$ which are associated
with a not null  probability $p_j$ must be in ${\cal B}^{(<)}$.

Finally, owing to the proof of the conjecture~(\ref{minentropy}) derived in this paper, one gets the simple expression
\begin{eqnarray}
C_{\chi }(\Phi^{\otimes m};E)&\leq &m \; \left[ S(\Phi(\rho_G^{(E)}))
- S(\Phi(|0\rangle\langle 0|)) \right]
\label{oneshotunassist1244}
\end{eqnarray}
and hence
\begin{eqnarray}
C(\Phi;E)&\leq &\; S(\Phi(\rho_G^{(E)}))
- S(\Phi(|0\rangle\langle 0|)) \;.
\label{oneshotunassist1233}
\end{eqnarray}
It turns out that for the channels~(\ref{def}) these bounds are attainable by exploiting  Gaussian encodings formed by Gaussian distribution of
coherent states~\cite{HOWE}, yielding the expressions given in Eqs.~(\ref{c1}), (\ref{c2}), (\ref{c3}), and (\ref{c4}).

\subsection{Mininal output entropy and classical capacity of the four fundamental Gaussian channels }

Exploiting the conjecture~(\ref{minentropy}), we simply have to write the explicit expression of $S(\Phi(|0\rangle\langle 0|))$ for each of the four fundamental classes of Gaussian bosonic channels. Together with the expression of $S(\Phi(\rho_G^{(E)}))$, we can use Eq.~(\ref{oneshotunassist1233}) to compute the classical capacity for each of the four fundamental channels (as explained, the upper bound is achieved with a Gaussian encoding). The minimal entropy and corresponding capacities are listed below.

\begin{itemize}
\item Thermal channel ${\cal E}_\eta^N$:
\begin{eqnarray}
S_{\min}^{(<)}[{\cal E}_\eta^N] &=& g((1-\eta)N) \;, \label{tg1}  \\
C({\cal E}_\eta^N;E) &=& g(\eta E + (1-\eta)N) - g((1-\eta)N)
\label{c1} \;;
\end{eqnarray}
\item Additive classical noise channel ${\cal N}_n$:
\begin{eqnarray}
S_{\min}^{(<)}({\cal N}_n) &=&g(n)\;,\\
C({\cal N}_n; E) &=&g(E+n)-g(n)\label{c2} \;;
\end{eqnarray}
\item Amplifier  channel ${\cal A}_\kappa^N$:
\begin{eqnarray}
S_{\min}^{(<)}[{\cal A}_\kappa^N] &=& g((\kappa -1)(N+1))
\;, \\ \nonumber
C({\cal A}_\kappa^N ;E)  \label{c3}
&=&g(\kappa E+(\kappa -1)(N+1)) \nonumber \\
&& -g((\kappa -1)(N+1))\;;
\end{eqnarray}
\item Contravariant amplifier  channel $\tilde{\cal A}_\kappa^N$:
\begin{eqnarray}
S_{\min}^{(<)}[\tilde{\cal A}_\kappa^N] &=& g(\kappa (N+1)-1)
\;, \\ \nonumber
C(\tilde{\cal A}_\kappa^N ;E)  \label{c4}
&=&g(\kappa N+(\kappa -1)(E+1)) \nonumber \\
&& -g(\kappa (N+1)-1)\;; \label{tg10}
\end{eqnarray}
\end{itemize}

In all these expressions, the function $g(x) = (x+1)\log_2 (x+1) - x \log_2 x$
refers to the von Neumann entropy of a Gibbs thermal Bosonic state with a mean photon number equal to $x$.

\subsection{Classical capacity of an arbitrary single-mode Gaussian channel}

Using the same approach,  an identity analogous to (\ref{minentropy})  can be shown to hold also for arbitrary   nondegenerate single-mode channels $\Psi$.
Indeed, as discussed in Refs.~\cite{cgh,hoc}, with few notable exceptions which will not be discussed here, all the single-mode channels can be expressed  as a proper concatenation of one of the maps $\Phi$ of Eq.~
(\ref{def}) together with two squeezing and/or displacement unitary transformations ${\cal U}$ and ${\cal V}$
acting  on the input and/or the  output of the communication line, i.e.
$\Psi = {\cal U}\circ \Phi \circ {\cal V}$ (the symbol  ``$\circ$" representing
concatenation of super-operators).
From Eq.~(\ref{minentropy}) it then follows that also  for these maps the min-entropy conjecture applies:
this time however  $S_{\min}^{(<)}[\Phi^{\otimes m}] $ is achieved over a Gaussian input obtained by properly squeezing the vacuum state
to compensate the action of ${\cal V}$.
As different from the case of the BGCs of Eq.~(\ref{def})
this however will guarantee the optimality of Gaussian inputs for $C(\Psi; E)$
and  the additivity of $C_\chi(\Psi; E)$  only  for large enough values
of the mean energy $E$ (see e.g. Ref.~\cite{LUPO}).

\subsection{Entanglement of Formation}

The two-mode Gaussian state $\rho(\kappa,N)$
resulting from applying a two-mode squeezing operation $U_\kappa$ to a
thermal state $\rho^{(N)}_G$, of covaraince matrix $(2N+1){\rm diag}(1,1)$
tensor the vacuum state (covariance matrix ${\rm diag}(1,1)$) reads
\begin{equation}
 \gamma=\left(\begin{array}{cc}
a I &  c\sigma_z  \\
c\sigma_z & b I  \\
\end{array} \right),
\end{equation}
where $\sigma_z={\rm diag}(1,-1)$, $a=2(N+1)\kappa-1$, $b=2(N+1)\kappa-(2N+1)$ and
$c=2(N+1)\sqrt{\kappa(\kappa-1)}$.

\paragraph{The upper bound}

It is easy to see that the covariance matrix can be decomposed as
\begin{equation}
\gamma=\gamma_{0}+M
\end{equation}
where
\begin{equation}
 \gamma_0=
 \left(\begin{array}{cc}
(2\kappa-1)I & 2\sqrt{\kappa(\kappa-1)}\sigma_z \\
2\sqrt{\kappa(\kappa-1)}\sigma_z & (2\kappa-1)I \\
\end{array} \right)
\end{equation}
is the covariance matrix of a two-mode squeezed vacuum state $\rho(\kappa,0)$, and
\begin{equation}
 M=2N
\left(\begin{array}{cc}
\kappa I & \sqrt{\kappa(\kappa-1)}\sigma_z \\
\sqrt{\kappa(\kappa-1)}\sigma_z & \kappa-1 I \\
\end{array} \right).
\end{equation}
The matrix $M$ can easily be shown to be positive, as it can be diagonalized
to ${\rm diag}(2\kappa-1,0)$. This implies that one can generate the Gaussian state
of covariance matrix $\gamma$ out of a two-mode squeezed vacuum state $\rho(\kappa,0)$ of
covariance matrix $\gamma_0$
and applying random correlated displacements (LOCC post-processing) to both modes.
This proves an upperbound to the entanglement of formation of such state,
as the optimal decomposition can only have lower cost in terms of entanglement, i.e.,
\begin{equation}
{\rm EoF}\left[\rho(\kappa,N)\right]\leq {\rm EoF}\left[\rho(\kappa,0)\right]=g(\kappa-1),
\end{equation}
where we used the fact that $\rho(\kappa,0)$ is a two-mode squeezed vacuum
state and pure.

\paragraph{The lower bound}
 To show that this quantity is also a lower bound for
 $\mbox{EoF}[\rho(\kappa,N)]$ one can use the fact  that  the reduced density matrix of $\rho(\kappa,N)$ coincides with
 the output state ${\cal A}_\kappa(\rho_G^{(N)})$
 of the quantum-limited amplifier  to the thermal state
 $\rho_G^{(N)}$. Therefore exploiting  the equivalence relation introduced
 in~\cite{WINTER} one has that
\begin{eqnarray}
\mbox{EoF}[\rho(\kappa,N)] &=& \min_{\{ p_j ; |\psi_j\rangle\}}
 \sum_j p_j S({\cal A}_\kappa(|\psi_j\rangle\langle\psi_j|)) \nonumber \\
 & \geq &
 S({\cal A}_\kappa(|0\rangle\langle0|)) \;,
 \end{eqnarray}
where the minimization being performed over the pure state
ensembles ${\cal E}=\{ p_j ; |\psi_j\rangle\}$ of  $\rho_G^{(N)}$
(i.e. $\sum_j p_j |\psi_j\rangle\langle \psi_j|=\rho_G^{(N)}$) and where the last inequality follows from
Eq.~(\ref{minentropy1}). Notice that since $\rho_G^{(N)}$ has finite
mean energy all the vectors $|\psi_j\rangle$ entering the ensemble
are elements of ${\cal B}^{(<)}$.

The combination of the upperbound and the lowerbound proves the equality
\begin{equation}
{\rm EoF}\left[\rho(\kappa,N)\right]={\rm EoF}\left[\rho(\kappa,0)\right]=
\kappa\log\left[\kappa\right]-(\kappa-1)\log\left[\kappa-1\right].
\end{equation}

\end{document}